# Glassy Aging Dynamics

P. Lunkenheimer, R. Wehn, U. Schneider, and A. Loidl

*Experimental Physics V, Center for Electronic Correlations and Magnetism, University of Augsburg, D-86135 Augsburg, Germany*

We present time-dependent dielectric loss data at different frequencies for a variety of glass formers after cooling below the glass temperature. The observed aging dynamics is described using a modified Kohlrausch-Williams-Watts law, which takes into account the time-dependent variation of the relaxation time during aging. It leads to values for the aging relaxation time and stretching exponent that are fully consistent with the results from equilibrium measurements performed at higher temperatures. Irrespective of the dynamic process prevailing in the investigated frequency region, the aging dynamics is always determined by the structural relaxation process.

The glass transition and the glassy state of matter belong to the most fascinating phenomena in condensed matter physics and are far from being really understood [1,2]. One of the hallmark features of glassy dynamics is the so-called "ergodicity breaking" [2], occurring not only in canonical glasses, but also, e.g., in spin glasses [3]. It arises when the sample is cooled too fast for the molecules to arrange into an equilibrated state, i.e. the sample "falls out of thermodynamic equilibrium". For typical cooling rates, this occurs close to the glass temperature $T_g$ and leaves the sample in a structural state corresponding to a higher temperature. In connection with this phenomenon, a variety of different, partly quite intriguing phenomena as aging, memory effects, and rejuvenation are observed. From a theoretical viewpoint, ergodicity breaking and the resulting non-equilibrium processes certainly are among the most challenging phenomena of glassy dynamics. They are also of considerable practical interest, e.g. for polymers, usually applied at temperatures not too far below $T_g$, where aging effects can lead to unwanted degradations of material properties. A famous approach for the description of non-equilibrium effects in glasses is the Tool-Narayanaswamy-Moynihan (TNM) formalism [4,5,6]. By introducing the concepts of a fictive temperature and a reduced time, it takes into account the so-called non-linearity of structural relaxation during aging, caused by the fact that the relaxation time itself is time-dependent during aging.

A straightforward experiment for the investigation of non-equilibrium effects in glass-formers is to monitor the time-dependent variation of physical quantities after quenching the sample below $T_g$. Then the so-called "physical aging" takes place, i.e. the physical quantities vary with time when the sample reapproaches equilibrium. This term was first used by Struick [7] to distinguish it from time-dependent processes involving chemical reactions [8]. There are numerous studies of physical aging in recent literature, however, while spin glasses [3] and to a smaller extent also polymers [7,9] have intensely been investigated, much fewer experiments were performed on canonical glass formers, so far. Especially, having in mind that dielectric spectroscopy has proven a key technique for the study of glassy dynamics in equilibrium, it is astonishing that investigations of glassy non-equilibrium dynamics in low molecular-weight glass formers with this technique are relatively scarce [5,10,11,12]. In the present letter, we provide detailed dielectric aging data on a variety of materials, belonging to different classes of glass formers, namely molecular glass formers and a glass-forming ionic melt, all having different fragilities [2,13,14], bonding types (hydrogen bridges or van der Waals bonds), and being characterized by an excess wing [15] or a well-developed Johari-Goldstein (JG) β-relaxation [16]. By introducing a new type of modeling of the experimental data, we demonstrate that the aging dynamics in all these materials is fully determined by the relaxation time and stretching parameter of the α-relaxation.

The dielectric permittivity of parallel plane capacitors was measured in the frequency range $10^{-4} \leq n \leq 10^6$ Hz using a Novocontrol alpha-analyzer and the autobalance bridge Hewlett-Packard HP4284. To keep the samples at a fixed temperature for up to nine weeks, a closed-cycle refrigerator system was used for temperatures below ambient and a home-made oven for temperatures above. The samples were cooled from a temperature 20 K above $T_g$ with the maximum possible cooling rate of about 3 K/min. The final temperature was reached without any temperature undershoot. As zero point of the aging times $t_{age}$, we took the time when the desired temperature was reached, typically about 100 s after passing $T_g$. The temperature was kept stable better than 0.1 K for all aging measurements.

Figure 1(a) shows the time dependence of the dielectric loss $e''$ of glycerol, detected at a frequency of 10 Hz. Glycerol, being a hydrogen-bonded network glass-former with $T_g \approx 185$ K and intermediate fragility $m \approx 53$ [14], belongs to the most investigated glasses formers. During aging, its loss decreases continuously with $t_{age}$ and finally becomes independent of time, indicating that the thermodynamic equilibrium state is reached after about $10^6$ s. To fit such kind of data, it seems natural to employ the time-honored Kohlrausch-Williams-Watts (KWW) law, which routinely is used to describe *equilibrium* relaxation processes in supercooled liquids [2]. Indeed, a perfect fit [solid line in Fig. 1(a)] is achieved using the ansatz

$$e''(t_{age}) = (e''_{st} - e''_{eq}) \exp\left[-\left(t_{age}/t_{age}\right)^{b_{age}}\right] + e''_{eq} \qquad (1)$$

where the indices "st" and "eq" indicate the values for $t_{age} \to 0$ and $\infty$, respectively, $t_{age}$ is the relaxation time, and $b_{age}$ the stretching parameter [12]. However, as was already revealed in the seminal study by Leheny and Nagel [12], both $t_{age}$ and $b_{age}$ do not agree with the corresponding parameters $t_a$ and $b_a$, determined from an extrapolation of fit results of the equilibrium spectra at $T > T_g$ and in addition they exhibit a considerable frequency dependence [17]. Especially, fits with eq. (1) yield $b_{age}$ always much smaller than the equilibrium value [12,17,18]. In the present case the extrapolated $b_a$ is 0.55 [19], significantly larger than $b_{age} = 0.29$ and it is not possible to fit the experimental data with $b_{age}$ fixed to 0.55 (dashed line). At first glance, such a behavior seems difficult to understand as it is reasonable that the variation of $\varepsilon''$ during aging can be traced back to structural rearrangements and thus should be governed by the same dynamics (i.e. relaxation time) and heterogeneity (i.e. stretching parameter) as the α-relaxation. This discrepancy is the manifestation of the non-linearity of sub-$T_g$ relaxation, a phenomenon addressed already long ago in the famous works by Tool and Narayanaswamy [4]. Namely it was pointed out that during aging the relaxation time $t_{age}$ itself is time dependent, i.e. subjected to aging. In the TNM formalism [4,5,6] this is taken into account by tracing back the aging-induced variation of physical quantities to the time-dependence of the so-called fictive temperature $T_f$ and by introducing an additional non-linearity parameter. This formalism (and other, mathematically nearly equivalent ones [11,20]) have been successfully used to describe various aging experiments (see, e.g., [6]), but its application is not straightforward, involving the numerical solution of a set of equations and requiring some assumptions for a proper evaluation. In the present work we analyze our data using a much simpler new approach as explained in the following.

Figure 1(c) shows $\varepsilon''(t_{age})$ for the shortest (100 s) and longest ($10^{6.5}$ s) aging time, together with spectra at $T > T_g$ [19]. Obviously, during aging the peak frequency $\nu_p$, which gives a good estimate of the relaxation rate $\nu_a = 1/(2\pi t_a)$, shifts towards lower values, which reflects the fact that the fictive temperature successively drops towards the actual temperature. Thus we use an ansatz for $\nu_a(t_{age})$, equivalent to eq. (1):

$$\nu_a(t_{age}) = 1/t_a = (\nu_{st} - \nu_{eq}) \exp\left[-\left(t_{age} 2\pi \nu_a\right)^{b_a}\right] + \nu_{eq} \quad (2)$$

This relation, $\nu_a = f(\nu_a)$, can be easily solved by iteration, resulting in a time dependence of $t_a$ that takes into account the non-linearity effects discussed above. Assuming that $t_{age} = t_a$ and $b_{age} = b_a$, the obtained $t_{age}(t_{age})$ is put into eq. (1), which then is used to fit the measured $\varepsilon''(t_{age})$. The resulting fit, with $b_{age} = b_a$ fixed to 0.55, is shown in Fig. 1(b) (solid line). The dashed and dash-dotted lines indicate conventional KWW curves using $b = 0.55$ and with $t = t_{st} = 1/(2\pi \nu_{st})$ and $t = t_{eq} = 1/(2\pi \nu_{eq})$, respectively, as obtained from the fit. It is the transition from the first to the latter curve during aging, which leads to the apparently larger "stretching" of the $\varepsilon''(t_{age})$ curve. Using this ansatz implies the assumption that the stretching remains unaffected by aging, i.e. that time-temperature superposition is valid. As the shift of the fictive temperature during aging is only few K for these experiments, this assumption is justified.

As shown in Fig. 2(a), this modified KWW ansatz was used to fit $\varepsilon''(t_{age})$ of glycerol for different frequencies leading to a perfect description of the experimental data. As of course the aging time dependence of $t$ must be independent of the frequency, the fits were performed simultaneously for all curves, with the parameters of eq. (2) identical for all $n$. The curves for different frequencies are only distinguished by two parameters, namely $e_{st}$ and $e_{eq}$ of eq. (1). Thus the number of parameters for each curve is four ($b$ being fixed to 0.55), however, two of them being common for all curves. This also implies that it should be possible to scale the $e''(t_{age})$ curves for different frequencies onto one master curve in the way given in Fig. 2(b), which indeed is the case. Further corroboration for the validity of the presented analysis arises from the values of the relaxation peak frequencies obtained from the fit parameters $n_{st}$ and $n_{eq}$, shown as open symbols in Fig. 1(c). They are fully consistent with a low-frequency extrapolation of the two aging curves at the shortest and longest aging times plotted in the figure.

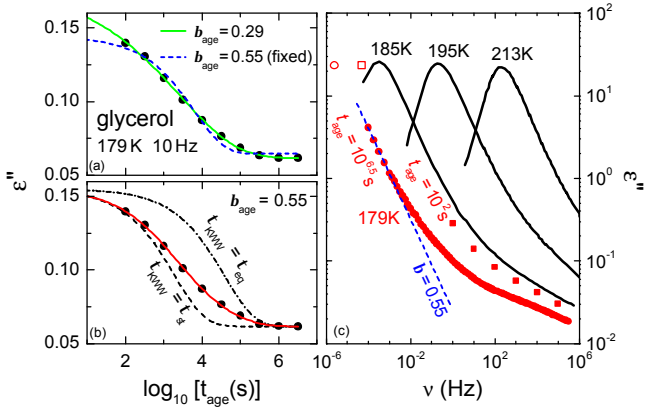

FIG. 1. (a) Aging-time dependent dielectric loss of glycerol. Lines: analysis with eq. (1) with all parameters free (solid) or with $b_{age}$ fixed at 0.55 (dashed). (b) Same data fitted with the modified KWW ansatz with $t_{age}(t_{age})$ determined from eq. (2) and $b_{age} = 0.55$ (solid line). The dashed and dash-dotted lines were calculated from eq. (1) with $t_{age} = t_{st}$ and $t_{age} = t_{eq}$, respectively. (c) Loss spectra for the shortest and longest $t_{age}$ (closed symbols), together with equilibrium data (solid lines) at higher temperatures [15,19]. The dashed line demonstrates a slope of 0.55. The open symbols indicate the peak frequencies for $t_{age} \to 0$ (square) and $\infty$ (circle), calculated from the parameters of the fit shown in (b).

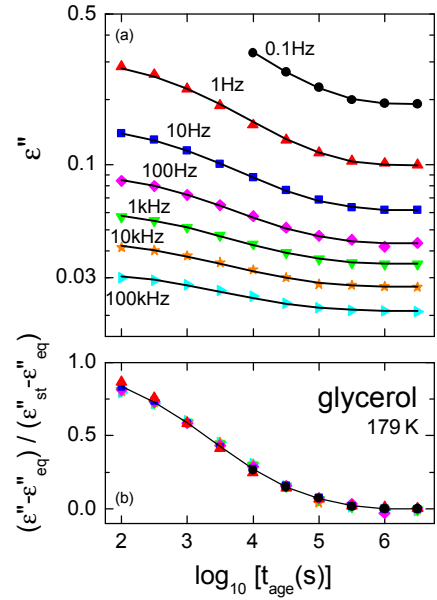

FIG. 2. (a) $\varepsilon''(t_{age})$ of glycerol at 179K for various frequencies. The lines are fits with the modified KWW ansatz with $b_{age}$ fixed to the equilibrium value of 0.55 and identical values of $\nu_{st}$ and $\nu_{eq}$ for all frequencies. (b) Scaling of the curves for different frequencies, using the values of $\varepsilon''_{st}$ and $\varepsilon''_{eq}$ obtained from the fits shown in (a).

relaxation peak [16]. Finally, $[Ca(NO_3)_2]_{0.4}[KNO_3]_{0.6}$ (CKN, $T_g \approx 333$ K, $m = 93$) is a typical ionic-melt glass former. For xylitol [Fig. 3(c)], remarkably the curve at 1 kHz crosses the one at 10 Hz. This can be ascribed to the low-frequency results reflecting the aging of the α-process, while at higher frequencies the aging in the regime of the JG $b$-process is observed [23,24,25]. For CKN [Fig. 3(d)], where the dielectric response is dominated by charge transport processes instead of dipolar reorientations, we plot the imaginary part of the dielectric modulus $M''$. In ionically conducting glass formers as CKN, it is common practice to consider this quantity instead of $e''$, because it usually reflects the structural relaxation behavior and because conductivity contributions, dominating $e''(n)$ at low frequencies, are suppressed in this representation [26,27]. As the relaxation peaks showing up in $M''(n)$ are known to be located at higher frequencies than those in $e''(n)$, also aging data at the low-frequency flank of the $a$-peak, *increasing* with time, could be collected [curve at 1 Hz in Fig. 3(d)].

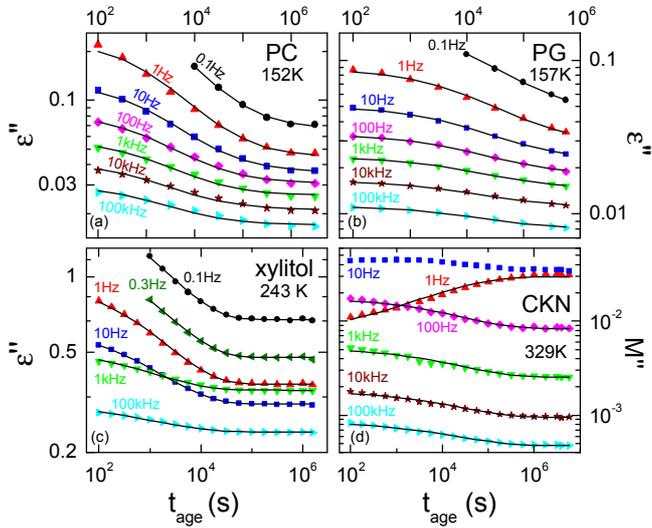

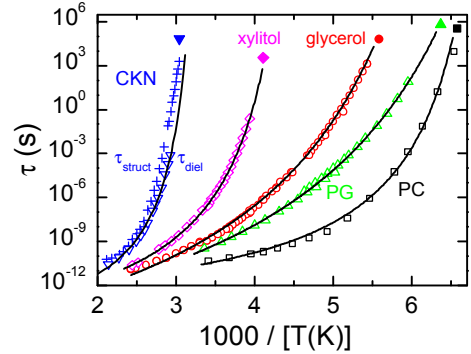

FIG. 3. $e''(t_{age})$ [(a)-(c)] respectively $M''(t_{age})$ (d) of four different glass formers for various frequencies. The lines are fits as described in Fig. 2(a) with $b_{age}$ = 0.6 (PC [15,19]), 0.58 (PG [22,25]), 0.43 (xylitol [25]), and 0.4 (CKN, from structural relaxation [27]).

FIG. 4. α-relaxation times from dielectric equilibrium measurements for the investigated glass formers (open symbols) [15,19,22,25,27]. The lines are fits with the Vogel-Fulcher-Tammann law. For CKN, in addition the structural relaxation time from mechanical spectroscopy is plotted (pluses) [28]. The closed symbols show the equilibrium average relaxation times $<t_{eq}>$, obtained from the fits in Figs. 2(a) and 3.

From Fig. 1(c), it becomes obvious that depending on frequency, the $e''(t_{age})$-curves in Fig. 2 reflect aging in the region of the right flank of the α-peak and in the region of the so-called excess wing showing up as a second power law at high frequencies [15,21]. Recent experiments have demonstrated that the excess wing is caused by a separate relaxation process [15,22,23]. Thus, independent of the dynamic process prevailing, the aging dynamics of $e''$ is always dominated by the α-process.

In Fig. 3, we show aging data for four further, different types of glass formers. Similar to glycerol, propylene carbonate (PC, $T_g \approx 159$ K, $m = 104$ [14]), propylene glycol (PG, $T_g \approx 168$ K, $m = 52$ [14]), and xylitol ($T_g \approx 248$ K, $m = 94$ [23,24]) are molecular glass formers. PC is a van-der-Waals liquid, while for PG and xylitol the formation of hydrogen-bonded networks can be assumed. While the equilibrium spectra of PC and PG exhibit an excess wing [15,21], xylitol [23,24] reveals a well pronounced JG β-

For all these cases, good fits of the spectra of Fig. 3 could be achieved using eqs. (1) and (2), the only exception being $M''(t_{age})$ of CKN at 10 Hz, exhibiting a shallow maximum. As for glycerol (Fig. 2), the parameters of eq. (1) were kept identical for all frequencies and $b_{age}$ was fixed to the equilibrium value. The resulting values of $t_{eq}$ are shown in Fig. 4 (closed symbols), together with $t_a(T)$ from equilibrium measurements at $T > T_g$ [15,19,22,25,27]. For all materials, $t_{eq}$ perfectly matches with the equilibrium curves. For ionically conducting glass formers as CKN, it is well established that decoupling phenomena become important at low temperatures. They lead to a deviation of the dielectric α-relaxation times $t_{diel}$, from the "true" structural α-relaxation times $t_{struct}$, determined, e.g., by mechanical experiments [27]. As becomes obvious from Fig. 4, $t_{eq}$ of CKN seems not to match $t_{diel}(T)$ (the curve would have to bend up unreasonably strong), but $t_{struct}(T)$, obtained from shear experiments [28]. This signifies the fact that the aging dynamics is governed by the structural relaxation time. Thus, despite of being determined by monitoring the aging of a quantity that is related to ionic transport (for CKN) or dielectric reorientation (for the other glass formers), the obtained $t_{eq}$ is a *structural α-relaxation* time. In the dipolar glass fomers, it matches the equilibrium dielectric relaxation times (Fig. 4) because translational-reorientational decoupling is less important or even absent (shown for glycerol in [29]).

An important result of this work is the finding that, independent of the spectral region (α-peak, excess wing, or β-peak) where the aging is monitored and independent of the relaxing entities (reorienting dipoles or ionic charge carriers), the aging dynamics of *ε″* is always dominated by the structural α-process. It is the structural rearrangement during aging, which in a direct way (by shifting the α-peak to lower frequencies) influences *ε″*($t_{age}$) in the α-peak region and in a more indirect way (by varying the structural "environment" felt by the relaxing entities) in the other regions. Here no statement needs to be made, in what way the aging may affect the other processes. For example, irrespective of the question if in the *β*-relaxation regime the aging affects mainly the *β*-relaxation time (i.e. leads to a frequency-shift of the peak) or reduces mainly the peak amplitude [15,16,30], in any case the time-dependence of this variation is determined by the structural α-relaxation. Only if during aging a transition between different regimes occurs, the situation must become more complicated, which is the case, e.g., for CKN at 10 Hz. Here during aging a transition from the low- to the high-frequency flank of the α-peak takes place (compare the non-equilibrium spectra at 325 K given in [27]), which leads to a maximum in *M″*($t_{age}$) [Fig. 3(d)] and thus the simple description with eqs. (1) and (2) must fail.

The straightforward description of aging dynamics with eqs. (1) and (2) successfully describes results at different frequencies with identical relaxation time and stretching parameter, both being fully consistent with equilibrium data. This ansatz works equally well also for the real part of the dielectric permittivity *ε′* (not shown) and it may also be used to describe other quantities. Our analysis reveals that all dynamic processes age in a similar way, determined by the structural α-relaxation dynamics. Success certainly justifies this to some extent phenomenological ansatz. However, one may ask, e.g., why eq. (2) should not be applied on $t_a$ instead of $n_a$. However, using such an ansatz does not lead to an equally consistent description, e.g. $t_{eq}$ does not match the $t_a(T)$ from equilibrium measurements. Obviously it is the rate that ages with KWW, but a theoretical foundation still needs to be found.

We thank C. A. Angell, R. Böhmer, J. C. Dyre, and R. Richert for illuminating discussions.